\documentclass[letterpaper]{jpconf}
\usepackage{graphicx}
\begin{document}

% Definitions
\def\nuc#1#2{${}^{#1}$#2}
\def\mee{$\langle m_{\beta\beta} \rangle$}
\def\mnu{$\langle m_{\nu} \rangle$}
\def\gnu{$\langle g_{\nu,\chi}\rangle$}
\def\mmod{$\| \langle m_{ee} \rangle \|$}
\def\mb{$\langle m_{\beta} \rangle$}
\def\BBz{$0\nu\beta\beta$}
\def\BBm{$0\nu\chi\beta\beta$}
\def\BBt{$2\nu\beta\beta$}
\def\BB{$\beta\beta$}
\def\bb{$\beta\beta$}
\def\bdec{\mbox{$\beta$ decay}}
\def\Mz{$|M_{0\nu}|$}
\def\ezero{$E_0$}
\def\Mt{$|M_{2\nu}|$}
\def\Tz{$T^{0\nu}_{1/2}$}
\def\Tt{$T^{2\nu}_{1/2}$}
\def\Tc{$T^{0\nu\,\chi}_{1/2}$}
\def\ms{$\Delta m_{\rm sol}^{2}$}
\def\ma{$\Delta m_{\rm atm}^{2}$}
\def\ts{$\theta_{\rm sol}$}
\def\ta{$\theta_{\rm atm}$}
\def\tot{$\theta_{13}$}
\def\ep{\mbox{e$^{+}$}}
\def\el{\mbox{e$^{-}$}}
\def\pos{\mbox{e$^{+}$}}
\def\belec{\mbox{$\beta$-electron}}
\def\bspec{\mbox{$\beta$ spectrum}}
\def\nue{\mbox{$\nu_{e}$}}
\def\numu{\mbox{$\nu_{\mu}$}}
\def\numub{\mbox{$\bar{\nu}_{\mu}$}}
\def\nub{\mbox{$\bar{\nu}$}}
\def\nueb{\mbox{$\bar{\nu}_{e}$}}
\def\nubx{\mbox{$\bar{\nu}_{x}$}}
\def\nutau{\mbox{$\nu_\tau$}}
\def\nui{\mbox{$\nu_1$}}
\def\nuii{\mbox{$\nu_2$}}
\def\nuiii{\mbox{$\nu_3$}}

\title{Introduction to the Double-Beta Decay Experimental Program}

\author{Steven R. Elliott}

\address{Los Alamos National Laboratory, Los Alamos, NM 87506, USA}

\ead{elliotts@lanl.gov}

\begin{abstract}
This document gives an overview of the technical issues and goals facing future double-beta decay experiments.
\end{abstract}

\section{Introduction}
The science of double-beta decay has been described in detail in a number of good review articles \cite{ELL02,ELL04,BAR04,AVI05}. In addition, these proceedings contain a lot of more current information on this rich
and exciting field. Previous experimental results are summarized in Ref.\cite{BAR06}, and the theoretical situation
concerning the matrix elements is summarized in Ref.\cite{SIM06}. Here we just list the very impressive experimental accomplishments to date in Table \ref{tab:ZeroNuResults}. 

In this article, we describe the issues facing the \BBz\ experimental program of the future. Table \ref{tab:ZeroNuProj} lists the active proposals for the future projects of which the author is aware. It presents an amazing variety of experimental techniques and expertise and is a tribute to the skill and ingenuity of the scientists involved. Many of these projects are well underway and many others have vigorous research programs, hence, the situation is extremely encouraging. While issues specific to a given project are described in other articles within these proceedings, there are numerous issues that are common to all these projects and it is these issues that are the focus of this article.

\begin{table}
\caption{\protect  A summary of the recent \BBz\ results. The \mee\ 
limits are those
deduced by the authors. All limits are at 90\% confidence level 
unless otherwise indicated. The columns providing the
exposure and background are based on arithmetic done by the author 
of this paper, who takes responsibility for any
errors in interpreting data from the original sources.}
\label{tab:ZeroNuResults}
\begin{center}
\begin{tabular}{lcclc}  \hline\hline
Isotope                    & Exposure                & Background  & Half-Life                     & \mee\    \\
                                  & (kmole-y)                &  (counts)        & Limit (y)                      & (meV)    \\ \hline
        $^{48}$Ca      & $5\times 10^{-5}$   &   0         & $>1.4 \times 10^{22}$       &  $<7200-44700$\cite{OGA04}               \\
        $^{76}$Ge      &   0.467                      &   21        & $>1.9 \times 10^{25}$       &  $<350$\cite{KLA01}              \\
        $^{76}$Ge      &   0.117                      &   3.5       & $>1.6 \times 10^{25}$       &  $<330-1350$\cite{AAL02}               \\
        $^{76}$Ge      &   0.943                      &   61        & $=1.2 \times 10^{25}$       &  $=440$\cite{KLA04}               \\
        $^{82}$Se      &   0.022                      &    7         & $>2.1 \times 10^{23}$         &  $<1200-3200$\cite{ARNO05,BAR06} \\
        $^{100}$Mo    & 0.131                       &    14         & $>5.8 \times 10^{23}$       &  $<600-2700$\cite{ARNO05, BAR06} \\
        $^{116}$Cd    & $1\times 10^{-3}$   & 14          & $>1.7 \times 10^{23}$       &  $<1700$\cite{DAN03}              \\
        $^{128}$Te     & Geochem.               &   NA        & $>7.7 \times 10^{24}$       &  $<1100-1500$\cite{BER93}              \\
        $^{130}$Te     & 0.07                          &   12         & $>2.4 \times 10^{24}$       &  $<400-1400$\cite{ARN05}              \\
        $^{136}$Xe     & $7\times 10^{-3}$   &   16        & $>4.4 \times 10^{23}$       &  $<1800-5200$\cite{LUE98}              \\
        $^{150}$Nd     & $6\times 10^{-5}$   &   0         & $>1.2 \times 10^{21}$       &  $<3000$\cite{DES97}              \\ \hline
\end{tabular}
\end{center}
\end{table}

\begin{table}
\caption{\protect  A summary of the \BBz\ proposals.}
\label{tab:ZeroNuProj}
\begin{center}
\begin{tabular}{lcl}  \hline\hline
 Collaboration                              & Isotope                        &      Detector   Description   \\ \hline
     CANDLES\cite{KIS04}          &    $^{48}$Ca               &        CaF$_{2}$ crystals in liq. scint.            \\
     COBRA\cite{ZUB01}             &     $^{116}$Cd            &       CdTe semiconductors             \\
     CUORE\cite{ARN04a}          &    $^{130}$Te             &     TeO$_{2}$ bolometers           \\
     DCBA\cite{ISH00}                 &    $^{82}$Se               &     Nd foils and tracking chambers             \\
     EXO\cite{DAN00}                  &    $^{136}$Xe             &     Xe TPC              \\
     GeH$_4$\cite{ROB05}         &    $^{76}$Ge             &      GeH$_4$ tracking ionization chamber  \\
     GEM\cite{ZDE01}                  &     $^{76}$Ge             &     Ge detectors in LN             \\
     GSO\cite{DAN01,WAN00}   &     $^{160}$Gd            &     Gd$_{2}$SiO$_{5}$ crystals in liq. scint.            \\
     Majorana\cite{GAI03}           &      $^{76}$Ge             &     Segmented Ge detectors            \\
     MOON\cite{EJI00}                 &      $^{100}$Mo           &     Mo foils and plastic scintillator             \\
     GERDA\cite{ABT04}             &     $^{76}$Ge              &       Ge detectors in LN             \\
     Nano-crystals\cite{MCD04} &                                      &      suspended nanoparticles             \\
     SeF$_6$\cite{NYG05}          &     $^{82}$Se              &      negative ion drifting SeF$_6$ TPC             \\
     Super-NEMO\cite{SAR00}   &     $^{82}$Se             &      foils with tracking             \\
     Xe\cite{CAC01}                      &     $^{136}$Xe           &    Xe dissolved in liq. scint.             \\
     XMASS\cite{MOR01}            &    $^{136}$Xe            &     liquid Xe             \\  \hline
\end{tabular}
\end{center}
\end{table}

\section{Producing a Result with Confidence}
The recent claim for positive evidence for \BBz\ in $^{76}$Ge\cite{KLA04a} has been controversial. One must ask why this rusult was not universally accepted and what types of evidence are required to ensure that future claims are embraced by the community.  Even though a peak is arguably present in the spectrum of this work, the signal was very weak and immersed in a large background. The background model had a fair amount of uncertainty including some unidentified lines near the region of interest. Supporting evidence to prove the peak was indeed due to \BBz\ and not some competing process was insufficient. Although future experiments will certainly have improved signal/background ratios, the supporting evidence question is more complicated. By noting that the physical process of \BBz\ has distinct characteristics, one can make a subjectively ordered list of potential supporting criteria.

\begin{itemize}
\item To show that \BBz\ likely exists, one needs a combination of:
  \begin{itemize}
  \item a clear peak at the correct \BBz\ energy value
  \item a demonstration that the event is a single-site energy deposit
  \item measured event distributions (spatial, temporal) are representative of \BBz
  \item a demonstration that the measured decay rate scales with isotope fraction
  \end{itemize}
\item To present a convincing case, one needs:
  \begin{itemize}
  \item an observation of the 2-electron nature of the \BBz\ event
  \item a demonstration that the  kinematic distributions (electron energy sharing, opening angle) match those of \BBz
  \item to observe the daughter nucleus appear in real time with the \BBz\ decay
  \item to observe the excited state decay process with parameters indicative of \BBz 
\end{itemize}
\item To remove all doubt, many of the above \BBz\ indicators should be:
  \begin{itemize}
  \item measured in several isotopes
  \end{itemize}
\end{itemize}

Although no experiment can demonstrate the entire list, the projects listed in Table \ref{tab:ZeroNuProj} all exploit a number of these.  

\section{Experimental Requirements}
Table \ref{tab:Signal} shows expected signal count rates in \BBz\ experiments as a function of neutrino mass. Present experiments are reporting half-life limits near $10^{25}$ years or $\approx$ 500 meV, whereas the next round of experiments hope to reach 100 meV. Such experiments should cover the degenerate mass region. Beyond that, experiments hoping to have sensitivity near the atmospheric mass scale will need about 1 ton of isotope. To obtain a signal-to-noise ratio of 1 will require a background level of $\approx$1 count/ton-year, which will be extremely challenging. Processes that are typically considered when estimating contributions to the background for \BBz\ include \BBt, naturally occurring radioactive isotopes, neutron-induced processes, and long-lived cosmogenic activities. 

For the current generation of experiments, energy resolutions are sufficient to prevent the tail of the \BBt\ energy spectrum from intruding into the \BBz\ peak region. Resolution will become a concern, however,  as we approach the ton scale.
Even so, resolution is a critical issue for the signal-to-noise ratio at any level of sensitivity. For example, an experiment with a factor 2 worse resolution will require a corresponding lower background for an equivalent ratio.

Naturally occurring radioactive materials, such as U and Th chain isotopes, occur as impurities in virtually all materials that make up an apparatus. The challenge is to ensure that the level of impurity is sufficiently low such that the decays of these isotopes won't mask the desired signal. The solution to this problem is mostly understood, but it is difficult to implement. Great progress has been made understanding materials and their associated U/Th contamination. Furthermore, purification and assay techniques have also improved. Even so, elaborate QA/QC programs will be required. In addition, to reach the ton scale, future purity levels will continue to greatly challenge assay capabilities.
Materials with purity levels of $\approx$1$\mu$Bq/kg or less will be required for ton scale experiments. It is difficult to assay materials to this level.
Hence, improvements are needed in the sensitivity of assay techniques such as mass spectroscopy, direct counting, and neutron activation analysis. Problems associated with long-lived cosmogenic isotopes are material dependent, but the problematic isotopes have been identified. Minimizing the surface exposure of detector materials and performing selected construction activities underground can mitigate much of this background contribution.

Unfortunately, neutron-induced backgrounds are more subtle. Neutrons originate from a number of sources. Those arising from ($\alpha$,n) and fission processes in a laboratory's surrounding rock have an energy up to $\approx$10 MeV and can be shielded effectively. Those arising from high-energy $\mu$ interactions in the rock and the detector shield materials can have very high energies and therefore are very penetrating. Unlike naturally-occurring radioactive isotopes, neutron-induced processes often don't have a unique signature that identifies the background process, which in turn provides clues to a mitigation plan. Instead, neutron related backgrounds are more likely to be a sum of a large number of processes, each of which is small by itself. This is especially true of (n,n'$\gamma$) reactions. To fully understand and plan for these backgrounds, the low-energy nuclear physics needs to be fully implemented into the simulation codes and verified. In some cases, the data required to do this doesn't yet exist. Moving to a deep site that shields the experiment from $\mu$'s will effectively reduce this background. Reference \cite{MEI06} estimates that a depth of $\approx$5000 m.w.e. will certainly suffice.

\begin{table}
\caption{\protect  A summary of the approximate \BBz\ signal rate for a number of 
neutrino masses. These estimates are for Ge, but are qualitatively similar
for most of the proposed isotopes.}
\label{tab:Signal}
\begin{center}
\begin{tabular}{cccc}  \hline\hline
 Neutrino Mass Scale   & \mee      &      Representative half life & Signal   \\ 
                                          & meV      &               years                        & counts/ton-year  \\ \hline
 Degenerate                   & 400        &               $10^{25}$                  & 530 \\
                                          &  100       &            $ 5 \times 10^{26}$    & 10 \\
 Atmospheric                  &  40         &            $ 5 \times 10^{27}$      & 1 \\
 Solar                              &     2          &          $ 10^{29}$                      & 0.05 \\  \hline
\end{tabular}
\end{center}
\end{table}

\begin{figure}[h]
\begin{center}
\includegraphics[width=24pc]{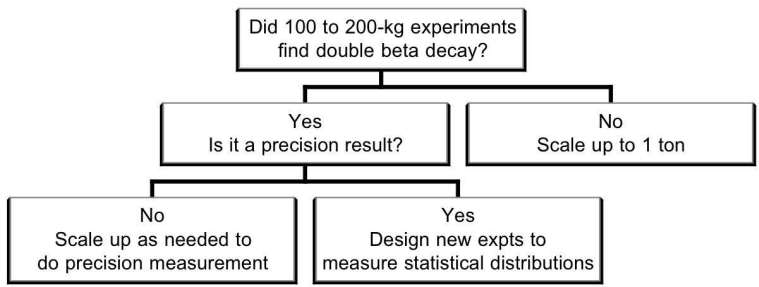}\hspace{2pc}%
\caption{\label{fig:DecisionTree} A cartoon of a decision tree for how the \BB\ program should proceed after the currently proposed generation of experiments.}
\end{center}
\end{figure}

Figure \ref{fig:DecisionTree} shows a flow chart indicating possible outcomes of future experiments and what they indicate for the future path of the overall \BBz\ program. After the current generation of experiments (100-200 kg) are complete, there will be a decision point regarding the following generation. If these experiments are null, then it will be necessary to build experiments with a ton of isotope to search for a signal at the atmospheric scale. Alternatively, if the experiments see a signal, there is different choice to make depending on the precision of the result. If the result is not a precision result ($\sim$10\%), then an expansion to the 1-ton scale is again warranted. Otherwise, if the result is a precision result, a follow-up experiment to measure the statistical distributions of kinematic parameters would be desired. Because the experimental design for measuring kinematic parameters may not be congruent with a simple scale up for a search experiment, planners will have to decide which direction to proceed after the current experiments are completed.

To reach a sensitivity at the solar scale, an experiment with 100 tons of isotope will be required. Such an experiment is not yet feasible for numerous reasons. Enrichment costs and production rates are not presently practical. Achieving the required excellent energy resolution (better than 1\%) in such a large experiment is also daunting. Schemes involving $10^6$ solid state detectors are conceivable, but costs would need to be greatly reduced to make that number of detectors affordable. Encouragingly, large multi-element detector electronics are improving and would not likely be a show-stopper. Alternatively, large volume detectors using metal-loaded liquid scintillator or Xe scale more easily and cost effectively. However, the energy resolution of such detectors is still too poor for this application. Significant research will be required on these technical difficulties if such an experiment is to be realized.

If \BBz\ is observed, we will want to extract all the underlying physics. The existence of the process would imply that neutrinos have a Majorana mass, but it doesn't necessarily mean that light neutrinos mediate the decay. Other possibilities include heavy neutrino exchange and R-pairity violating super-symmetry particle exchange. See Ref. \cite{ELL04} and references therein for a discussion of alternative decay mechanisms. The matrix elements, however, are different for differing processes and this leads to a possible technique for isolating the dominate one. By comparing \BBz\ rates in several isotopes one might be able to identify the underlying physics. If one uses the present theoretical values for the matrix elements as a guide, it appears that  3 or more experiments, each with a total uncertainty (theory, statistical, systematic) of less than about 40\% will be required.

\section{The Majorana Phases}
The linear combination of the neutrino mass eigenstates ($\nu_k$) that mix to form the weak interaction eigenstates 
($\nu_{\alpha}$) is given by a mixing matrix, $U_{\alpha k}$. This matrix may contain as many as 3 physically meaningful phases, two
of which (the Majorana phases $\alpha_{21}, \alpha_{31}$) contribute to the effective double-beta decay mass (\mee). These phases do not contribute to the effective beta decay mass (\mb) or the differences in the squares of the neutrino masses (\ms) as measured in oscillation experiments. 

\begin{figure}[h]
\begin{minipage}{18pc}
\includegraphics[width=15pc]{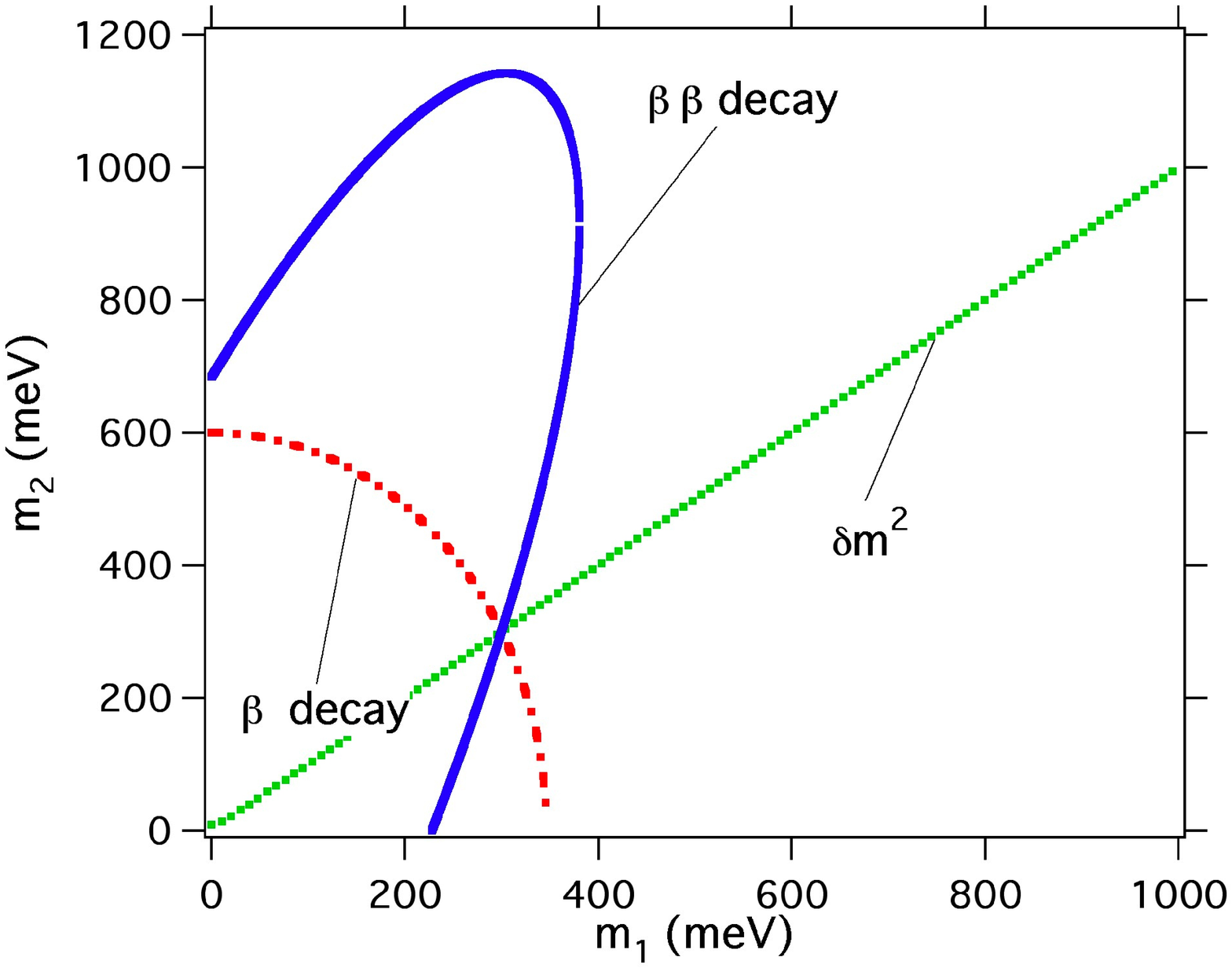}
\caption{\label{fig:m1m2plot}A consistency plot for the neutrino mass eigenvalues $m_1$ and $m_2$, for various hypothetical measurements. This set of curves indicates how measured values of $\Sigma$, \mee, \ms, and \mb\ constrain the mass eigenvalues. See text for a description of the chosen input parameters.}
\end{minipage}\hspace{2pc}%
\begin{minipage}{18pc}
\includegraphics[width=12pc]{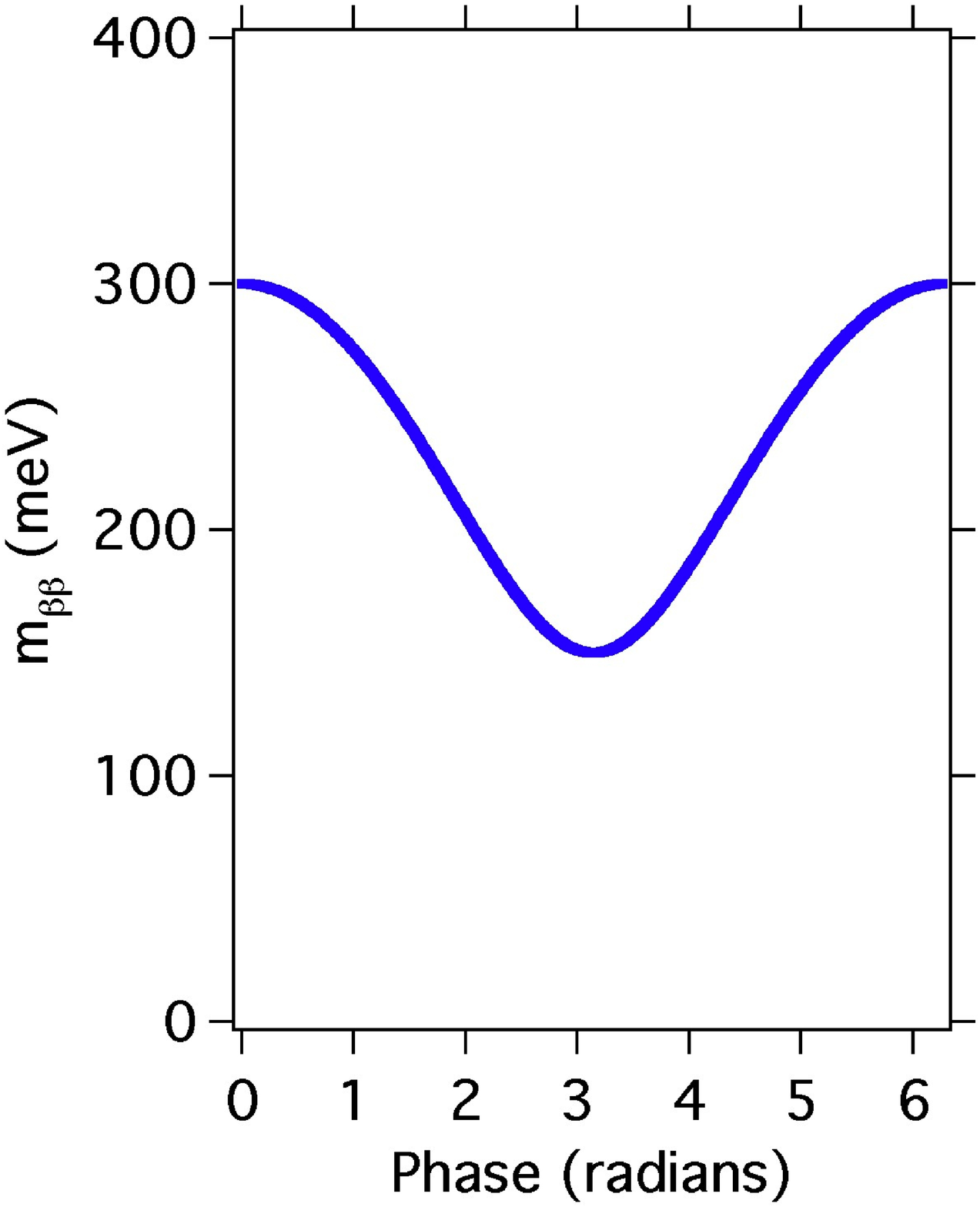}
\caption{\label{fig:phase}For the same parameters as those in Fig. \ref{fig:m1m2plot}, a plot of \mee\ as a function of the Majorana phase.}
\end{minipage} 
\end{figure}

If $U_{e3} \neq 0$, then both phases contribute to \mee\ and since no other experiment has been identified that is sensitive to the phases, there will be an ambiguity in trying to extract the phase values. You can't deduce two parameters from one data point.  However,  if $U_{e3} = 0$, only one of the Majorana phases contribute to \mee\ and it could, in principle be extracted. To show this we compare measurements of \mee, \mb, and \ms\ for a toy model. Figure \ref{fig:m1m2plot} was drawn for $m_1 = 300$ meV, \ms\ =(9 meV)$^2$, $U_{e1}=0.866$, $U_{e2}=0.5$, and $\alpha_{21}$ = 2.5 radians. These result in \mb\ = 300 meV and \mee = 171 meV. Note that the 3 measured parameters are all plotted as functions of the mass eigenvalues and they agree at only one point and then only if the correct value for $\alpha_{21}$ is chosen. 

However, to determine the value of the phase with any precision requires great accuracy on \mee. Figure \ref{fig:phase} shows \mee\ as a function of the value of the phase for the above toy model. Its clear that if the uncertainty on \mee\ is 50\%, no information regarding $\alpha$ is obtained. For a useful determination of the phase, even in this simplistic two-flavor model, a precision nearing 10\% is required. Note that a similar analysis can be found in Reference \cite{BAR02} and a three-neutrino-species example is presented in Reference \cite{ELL04}.

\section{Conclusions}
This is a very exciting time for \BB. The experimental program is poised to make great strides into a region that will greatly impact neutrino physics. Since neutrino oscillations imply that neutrinos have mass, even null \BBz\ experiments will constrain the possible neutrino mass spectra. If one sees \BBz\ in the upcoming experiments, the qualitative physics conclusion will be extremely exciting. However, to fully understand all the underlying physics, precision measurements in several nuclei will be needed.

\ack{Acknowledgments}
This work was supported in part by Laboratory Directed Research and Development at LANL. I thank Peter Doe and Petr Vogel for careful readings of this manuscript.


\section{References}

\begin{thebibliography}{99}
\bibitem{ELL02} Elliott Steven R, and Petr Vogel, 2002 {\it Annu. rev. Nucl. Part. Sci.} {\bf 52} 115 
\bibitem{ELL04} Elliott S R and J Engel, 2004 {\it J. Phys.} {\bf G30} R183
\bibitem{BAR04} Barabash A S, 2004 {\it Physics of Atomic Nuclei} {\bf 67}, No. 3, 438 
\bibitem{AVI05} F.T Avignone III, G.S. King III and Yuri Zdesenko, 2005 {\it New Journal of Physics} {\bf 7} 6

\bibitem{BAR06}Barabash A S, "Double beta decay experiments:
Past and present results", presented at Neutrino 2006 

\bibitem{SIM06} Simkovic Fedor 2006, "The nuclear matrix elements for double-beta decay ", presented at Neutrino 2006

\bibitem{OGA04} Ogawa I \etal  2004 {\it Nucl. Phys.} A {\bf 730} 215; S. Umehara, "CANDLES for double beta decay of $^{48}$Ca", presented
at TAUP 2005 %CANDLES
\bibitem{KLA01} Klapdor-Kleingrothaus H V, P\"{a}s H and Smirnov A Yu 2001 {\it Phys. Rev. D} {\bf 63} 073005
\bibitem{AAL02} Aalseth C E \etal  2002 {\it Phys. Rev.} D {\bf 65} 092007
\bibitem{DAN03} Danevich F A \etal  2003 {\it Phys. Rev.} C {\bf 68} 035501
\bibitem{BER93} Bernatowicz T \etal  1993 {\it Phys. Rev.} C {\bf 47} 806
\bibitem{KLA04} Klapdor-Kleingrothaus H V, Dietz A, Krivosheina I V and  O. Chkvorets 2004 {\it Phys. Lett.} B {\bf 586} 198; {\it Nucl. Instrum. Meth.} A {\bf 522} 371
\bibitem{ARNO05} Arnold R \etal 2005 {\it Phys. Rev. Lett.} {\bf 95} 182302 %NEMO-3
\bibitem{ARN05} Arnaboldi C \etal  2005 {\it Phys. Rev. Lett.} {\bf 95} 142501; R. Maruyama, "Cuore and Cuoricino:
A bolometric search for neutrinoless double beta decay", presented at Neutrino 2006 %cuoricino
\bibitem{LUE98} Luescher R \etal  1998 {\it Phys. Lett.} B {\bf 434} 407
\bibitem{DES97} De Silva A \etal  1997 {\it Phys. Rev.} C {\bf 56} 2451


\bibitem{KIS04} Kishimoto T 2004, private communication %CANDLES
\bibitem{ZUB01} Zuber K 2001 {\it Phys. Lett.} B {\bf 519} 1; J. Wilson, "The COBRE experiment", presented at Neutrino 2006  %COBRE
\bibitem{ARN04a} Arnaboldi C \etal  2004a {\it Nucl. Instrum. Meth.} {\bf A518} 775 %CUORE
\bibitem{ISH00} Ishihara N \etal  2000 {\it Nucl. Instrum. Meth.} A {\bf 443} 101 %DCBA
\bibitem{DAN00} Danilov M \etal  2000 {\it Phys. Rev.} C {\bf 62} 044501; A. Piepke, "New techniques: EXO, MOON, SuperNEMO", presented at Neutrino 2006 % EXO
\bibitem{ROB05} Robertson RGH 2005, private communication % germane TPC
\bibitem{ZDE01} Zdesenko Yu G, Ponkratenko O A and Tretyak V I 2001 {\it J. Phys.} {\bf G27} 2129 %GEM
\bibitem{DAN01} Danevich F A \etal  2001 {\it Nucl. Phys.} A {\bf 694} 375 %GSO
\bibitem{WAN00} Wang S C, Wong H T and Fujiwara M 2000 {\it Nucl. Instrum. Meth.} A {\bf 479} 498-510 %GSO
\bibitem{GAI03} Gaitskell R \etal  2003 {\it Preprint} nucl-ex/0311013 %Majorana
\bibitem{EJI00} Ejiri H \etal  2000 {Phys. Rev. Lett.} {\bf 85} 2917; H. Nakamura, "MOON for double beta decay experiment and MOON-1 prototype detector status", presented at TAUP 2005 %MOON
\bibitem{ABT04} Abt I \etal  2004 LNGS-LOI 35/04, {\it Preprint} hep-ex/0404039; S. Sch\"{o}enert, "New techniques in $0\nu\beta\beta$ germanium
experiments", presented at Neutrino 2006 % MPI naked Ge
\bibitem{MCD04} McDonald A 2004 private communication for members of the SNO Collaboration %nanocrystals
\bibitem{NYG05} Nygren D 2005, private communication
\bibitem{SAR00} Sarazin X \etal  2000 {\it Preprint} hep-ex/0006031
\bibitem{CAC01} Caccianiga B and Giammarchi M G 2001 {\it Astropart. Phys.} {\bf 14} 15 %Xe in CTF
\bibitem{MOR01} Moriyama S \etal  2001 presented at XENON01 Workshop, Dec., Tokyo, Japan

\bibitem{KLA04a}Klapdor-Kleingrothaus H V, A Dietz, I V Krivosheina and O Chkvorets, 2004 {\it Nucl. Instrum. Meth.} {\bf A522} 371 

\bibitem{MEI06} Mei D-M, and A Hime 2006 {\it Phys. Rev.} {\bf D73} 053004 

\bibitem{BAR02}Barger V, S L Glashow, P Langacker, D Marfatia 2002 {\it Phys.Lett.} {\bf B540} 247% no-go theorem for MJ phases
\end{thebibliography}
\end{document}